# Layer-dependent surface potential of phosphorene and anisotropic/layer-dependent charge transfer in phosphorene-gold hybrid system

Renjing Xu,[1] Jiong Yang,[1] Yi Zhu,[1] Yan Han,[1] Jiajie Pei,[1,2] Ye Win Myint,[1] Shuang Zhang,[1] and Yuerui Lu[1*]

[1]Research School of Engineering, College of Engineering and Computer Science, the Australian National University, Canberra, ACT, 0200, Australia

[2]School of Mechanical Engineering, Beijing Institute of Technology, Beijing, 100081, China

* To whom correspondence should be addressed: Yuerui Lu (yuerui.lu@anu.edu.au)

**ABSTRACT:** The surface potential and the efficiency of interfacial charge transfer are extremely important for designing future semiconductor devices based on the emerging two-dimensional (2D) phosphorene. Here, we directly measured the strongly layer-dependent surface potential of mono- and few-layer phosphorene on gold, which confirms with the reported theoretical prediction. At the same time, we used an optical way - photoluminescence (PL) spectroscopy to probe the charge transfer in phosphorene-gold hybrid system. We firstly observed highly anisotropic and layer-dependent PL quenching in the phosphorene-gold hybrid system, which is attributed to the highly anisotropic/layer-dependent interfacial charge transfer.

Phosphorene, a new family member of two-dimensional (2D) layered materials, has attracted tremendous attention owing to its unique anisotropic manner[1-6], layer-dependent photoluminescence[7, 8], and quasi-one-dimensional (1D) excitonic nature[9, 10], which are all in drastic contrast with the properties of other 2D materials, such as graphene[11] and transition metal dichalcogenide (TMD) semiconductors[12-15]. Particularly, phosphorene with narrow band gaps ranging from mid-infrared to near-infrared wavelengths can fill the space between the gapless graphene and the comparably large gap TMD semiconductors, leading to many new electronic and optoelectronic devices[1-10], including field effect transistors, sensors, light emitting diodes, solar cells and so on. All these devices involve the photon-to-electron or electron-to-photon conversions, in which the interfacial charge transfer plays an important factor[16-18]. Up to date, the important properties of metal-phosphorene interfaces are still underexplored, including the direct measurement of the surface potential of phosphorene, the interfacial charge transfer in the phosphorene-metal hybrid system, etc.

**RESULTS AND DISCUSSION**

In this work, we directly measured the surface potential of mono- and few-layer phosphorene on gold substrate by Kelvin probe force microscope (KPFM). The measured surface potential shows strong layer dependence, which is consistent with reported theoretical prediction[19]. At the same time, we used photoluminescence (PL) spectroscopy to probe the charge transfer in phosphorene-gold hybrid system. We firstly observed highly anisotropic and layer-dependent PL quenching, which is attributed to the highly anisotropic/layer-dependent interfacial charge transfer in the phosphorene-gold hybrid system. Our results provide very useful information for heterostructure device design and pave new avenues for future new optoelectronic devices using phosphorene.

Surface potential, also known as work function, is very important for semiconductor devices, since it determines the interfacial energy barrier and thus the efficiency of carrier injections in electronic and optoelectronic devices. Here, a powerful tool KPFM was used to directly measure the surface potential of mono- and few-layer phosphorene flakes on gold substrates (Figure 1). In experiments, mono- and few-layer phosphorene flakes were mechanical exfoliated onto gold substrates. The thin layers were firstly identified by the contrast under optical microscope (Figure 1b) and the layer numbers were then quickly and precisely determined by our recently developed technique phase-shifting interferometer[20] (Figure 1c and 1d, Figure S1). The layer number was also confirmed by the measured PL spectra[20]. For the surface potential measurements, we applied an AC modulation voltage and a DC bias voltage between the sample and the tip (Figure 1a). The resulting electrostatic forces acting on the tip were detected by using a lock-in technique to analyze the amplitude of the tip's oscillation in the KPFM system. The amplitude signal includes the DC bias voltage plus the work function difference between the tip and the sample. The compensated DC voltage signal, controlled by a feedback circuit, corresponds to the local surface potential of the sample surface[21, 22]. The resulting KPFM image (Figure 1e) maps the variation of the surface potential. From this KPFM image, we could determine the surface potential difference between the gold substrate and the phosphorene flake, defined as $\phi_{Au} - \phi_{BP}$, where $\phi_{Au}$ and $\phi_{BP}$ are the work functions of gold and phosphorene flake, respectively. The measured statistical surface potential differences for mono-layer (1L) to five-layer (5L) phosphorene and thick flakes (relative to gold surface) are 514 ± 15, 536 ± 6, 555 ± 7, 599 ± 14, 665 ± 21 and 693 ± 10 meV, respectively, which is consistent with the trend of previous simulated results[19]. The KPFM measurements were all completed within 40 minutes after phosphorene exfoliation to minimize the influence from sample degradation.

The layer-dependent surface potential of the phosphorene can enable the engineering of the exciton dynamics in the phosphorene-metal heterostructures, including generation, dissociation, transfer, and recombination, which bears tremendous significance for fundamental and applied interests[18, 23]. Here, we study charge transfer dynamics in phsophorene-gold hybrid system using PL spectroscopy (Figure 2), which has been used to probe the charge transfer in TMD 2D semiconductor heterostructures[16-18, 24]. In our experiment, few-layer phosphorene flakes were mechanically exfoliated and transferred onto the edge of a gold pad, which was prepatterned onto a $SiO_2$/Si (275 nm thermal oxide substrate) substrate. The transfer process was carefully carried out to make sure half of the few-layer phosphorene flake lays on the $SiO_2$/Si substrate whilst the other half is on the gold (Figure 2a and 2b), which will give us precise comparison of the PL emissions from these two halves. From the optical microscope image (Figure 2b), the first half of the few-layer phosphorene sample on the $SiO_2$/Si substrate presents much larger contrast than the other half on the gold, which is because the thickness of the $SiO_2$ was pre-designed for the best contrast[25]. The layer number was precisely determined by phase-shift interferometry (PSI) (Figure 2c & 2d, Figure S2 & S3), which has been demonstrated to be a fast and precise way to identify the layer number of phosphorene in our previous report[26]. The layer number identification was also confirmed by comparing the measured PL spectra with our previous reports[8, 26]. Figure 2e shows the measured PL spectra of the bilayer phosphorene samples on both gold and $SiO_2$/Si substrate, using a 532 nm Nd:YAG laser as the excitation source. The PL intensity of the main peak at 945 nm from the 2L phosphorene on Au is much lower than that from the second half of 2L phosphorene on $SiO_2$/Si substrate, which means the Au substrate leads to significantly PL quenching. This PL quenching is due to the fast interfacial charge transfer between the gold and the 2L phosphorene sample. From the energy diagram shown in Figure 2f, the photo-excited holes are attracted towards the Au-phosphorene interface and then annihilated by the electrons from the gold[27]. Hence, as the consequence of the fast charge transfer, the radioactive recombination will be eliminated, leading to the PL quenching in the phosphorene-gold hybrid system.

In order to fabricate high-efficiency solar cells based on phosphorene, it is very important to optimize the charge transfer efficiency and thus the carrier collection efficiency in the phosphorene-metal hybrid system[18, 28]. Owing to the unique puckered structure, phosphorene has been demonstrated to show highly anisotropic properties, such as highly anisotropic excitons[9, 29], carrier mobility[30], carrier effective masses[17, 30], thermal expansion coefficient[31] and shear factor[31], which are all in drastic contrast with the properties of graphene[11] and TMD semiconductors[12-14]. Therefore, an anisotropic charge transfer in the phosphorene-gold hybrid system could be expected, which makes it more critical to precisely determine the right orientation for the optimized charge transfer. Here, we used the angle-resolved PL quenching measurements to characterize the anisotropic charge transfer in the phosphorene-gold hybrid system. In the setup of the angle-resolved PL measurement, a linearly polarized Nd:YAG laser with a wavelength of 532 nm was used as the excitation source. The polarization angle of the incident light is controlled by an angle-variable half-wave plate. The polarization angle of the PL emission (θ) is characterized by inserting an angle-variable polarizer in front of the detector. We firstly used the excitation polarization dependent PL measurements to determine the crystalline orientation of the 2L phosphorene flake on $SiO_2$/Si substrate[10]. Then the armchair direction is selected as the zero-degree reference for θ (Figure 3a, inset). Next, we measured the polarization dependence of the PL emission by fixing the excitation polarization along the armchair direction. Figure 3a presents the polarization dependent PL emission from the first half of the 2L phosphorene flake on $SiO_2$/Si substrate (Figure 2b). The red fitting curve shows that the PL intensity follows sinusoid with period of $\sim 180^o$. The PL intensity reaches its maximum at $\theta = 0^o$ (armchair direction) and its minimum at $\theta = 90^o$ (zigzag direction), which is consistent with our previous report[10]. The measured PL emission from the second half of the 2L phophorene flake on Au (Figure 3b) is also linearly polarized, whereas its maximum PL intensity appears at $\theta = 50^o$ instead of $0^o$. We attribute this angle difference to be caused by the anisotropic interfacial charge transfer between the phosphorene and gold. Here we use a quenching factor, which is defined by $\frac{\text{PL intensity on } SiO_2/Si}{\text{PL intensity on Au}}$, to characterize the charge transfer. Figure 3c shows the measured quenching factor as a function of the emission polarization angle. The quenching factor is strongly dependent on the polarization angle, which can be fitted by a sinusoidal curve with a period of $\sim 180^o$. The measured maximum quenching factor was 6, which occurs at $\theta = 0^o$, while the minimum quenching factor is 1.5, which occurs at $\theta = 90^o$. This strongly anisotropic PL quenching can be understood by the anisotropic interfacial charge transfer between the phosphorene and gold. Since the PL quenching effect between the semiconductor materials and metal is basically due to the transport of photo-excited charges[24, 27]. Phosphorene has the maximum charge mobility along the armchair direction[32], which leads to the maximum speed of charge transfer and thus the maximum quenching factor along the armchair direction.

Besides the charge mobility discussed above, the Schottky energy barrier is another important factor to determine the efficiency of interfacial charge transfer[17, 18]. Previous calculations have predicted that phosphorene owns layer-dependent band alignment and there is a significant shift of valence band edge upon thickness variation, which can be useful for tuning Schottky barrier to promote hole injection efficiency[33]. This prediction is confirmed from our following layer-dependent PL quenching experiments in the phosphorene-gold hybrid system. The measured quenching factor of a 3L phosphorene on Au is around one order of magnitude higher than that of a 2L phosphorene, for emission polarization of both zigzag and armchair directions (Figure 4a), respectively. This strongly layer-dependent PL quenching in phosphorene is in great contrast with that in $MoS_2$, where the quenching factor does not make large changes for mono- and few-layer $MoS_2$[24]. The layer-dependent PL quenching in phosphorene-gold hybrid system can be understood from their band diagrams, as shown in Figure 4b. The band diagrams were generated based on the calculated valence band maximum (VBM) and conduction band minimum (CBM) for few-layer phosphorene[19] and our measured surface potential difference between phosphorene and Au. As shown in Figure 1f, the measured surface potential difference between 3L phsophorene and gold (~555 meV) is larger than that for 2L phosphorene (~536 meV), which results in larger Schottky barrier for 3L phosphorene-Au hybrid system than that for 2L phosphorene (Figure 4b). This larger Schottky barrier will increase the force to pull holes towards the phosphorene-gold interface, leading to more efficient interfacial charge transfer and thus much larger PL quenching for 3L than 2L phosphorene in the phosphorene-gold hybrid system.

**CONCLUSIONS**

In conclusion, we directly measured the layer-dependent surface potential of mono- and few-layer phosphorene on gold substrate, which is consistent with reported theoretical prediction[19]. At the same time, we used an optical way - PL spectroscopy to probe the charge transfer in phosphorene-gold hybrid system. We firstly observed highly anisotropic and layer-dependent PL quenching, which is attributed to the highly anisotropic/layer-dependent interfacial charge transfer in the phosphorene-gold hybrid system. Our results open the door for **designing** future electronic and optoelectronic heterostructure devices using phosphorene.

**EXPERIMENTAL METHODS**

We used mechanical exfoliation to transfer[34] a phosphorene flake onto the edge a pre-patterned gold electrode on a $SiO_2$/Si substrate (275 nm thermal oxide on $n^+$-doped silicon). The gold electrodes were patterned by conventional photolithography, metal deposition, and lift-off processes. All PL and polarization measurements were conducted using a T64000 micro-Raman system equipped with a charge-coupled device (CCD) and InGaAs detectors, along with a 532nm Nd:YAG laser as the excitation source. Subsequent to PSI measurement, the sample was placed into a Linkam THMS 600 chamber, with a slow flow of nitrogen gas to prevent degradation of the sample[8]. To avoid laser-induced sample damage, all PL spectra were recorded at low power levels: P ~ 20 μW. The surface potential was measured by KPFM. The model of the KPFM used in the experiments was Bruker Multimode VIII with Peak Force TUNA.

**Acknowledgements**

We would like to thank Professor Chennupati Jagadish and Professor Barry Luther-Davies and from The Australian National University, for their facility support. We acknowledge financial support from ANU PhD student scholarship, China Scholarship Council, Australian Research Council and ANU Major Equipment Committee.

**Competing financial interests**

The authors declare that they have no competing financial interests.

**Supporting Information Available:** Details of theoretical calculation methods, more experimental measurement results and data analysis, are shown in Supporting Information.

## FIGURE CAPTIONS

**Figure 1 | Measured surface potential of mono- and few-layer phosphorene by Kelvin probe force microscope (KPFM). a**, Schematic plot of KPFM measurement. **b**, Optical microscope image of the sample with two- and three-layer (2L and 3L) phosphorene on gold substrate. **c**, Phase-shifting interferometry (PSI) image of the 2L and 3L phosphorene from the box enclosed by the dash line in **(b)**. **d**, PSI measured optical path length (OPL) values versus scan position for 2L and 3L phosphorene (or black phosphorous, shorted as BP) along the dash line in **(c)**. The measured optical path length (OPL) values of the sample on gold are 6.2 and 9.4 nm, indicating bi- and tri-layer, respectively (Figure S1). **e**, Surface potential contour image of the dashed line box region in **(b)**, measured by KPFM. The surface potential differences between the gold substrate and the phosphorene layers (2L and 3L) are also shown, respectively. The surface potential difference between the gold substrate and the phosphorene flake is defined as $\phi_{Au} - \phi_{BP}$, where $\phi_{Au}$ and $\phi_{BP}$ are the work functions of gold and phosphorene flake, respectively. **f**, Measured layer-dependent surface potential differences between the gold substrate and phosphorene flakes. The error bar represents the statistical measurement variations from at least two samples for each layer number.

**Figure 2 | Photoluminescence (PL) quenching in phosphorene-gold hybrid system. a**, Schematic plot of the device structure. A few-layer phosphorene flake was mechanically exfoliated and dryly transferred onto the edge of a gold pad, which was pre-patterned onto a SiO$_2$/Si (275 nm thermal oxide substrate) substrate. The transfer process was carefully carried out to make sure half of the few-layer phosphorene flake lays on the SiO$_2$/Si substrate whilst the other half is on the gold, which will give us precise comparison of the PL emissions from

these two regions. **b**, Optical microscope image of the mechanically transferred few-layer phosphorene flakes, with half on gold and half on SiO$_2$. **c**, PSI image of phosphorene samples from the lower box on SiO$_2$/Si substrate enclosed by the dash line in (**b**). **d**, PSI measured OPL values versus scan position for the phosphorene along the dash line in (**c**). **e**, Measured PL spectra from the 2L phosphorene on Au and SiO$_2$/Si substrate, with the background PL spectra for comparison. **f**, Schematic plot of the band diagram, showing the mechanism of the PL quenching in phosphorene-gold hybrid system.

**Figure 3 | Anisotropic PL quenching/charge transfer in phosphorene-gold hybrid system. a-b**, Measured PL peak intensities as a function of the emission polarization angle θ, from the 2L phosphorene sample half on SiO$_2$/Si (**a**) and the second half on gold (**b**), respectively. The red lines are the fitted sinusoidal curves. Inset: Schematic plot of the top view of phosphorene lattice structure and coordinates for polarization angle θ. **c**, The measured PL quenching factor from the 2L phosphorene sample, as a function of the emission polarization angle. Quenching factor is defined by $\frac{\text{PL intensity on SiO}_2/\text{Si}}{\text{PL intensity on Au}}$. The red line is the fitted sinusoidal curve.

**Figure 4 | Layer-dependent PL quenching/charge transfer in phosphorene-gold hybrid system. a**, Measured PL quenching factors from 2L and 3L phosphorene samples, with emission polarization angles along the armchair and zigzag directions, respectively. The error bar represents the statistical measurement variation from at least two samples. **b**, Schematic plot of the band diagrams for 2L and 3L phosphorene-gold hybrid systems. The band diagrams were generated based on the calculated valence band maximum (VBM) and conduction band minimum (CBM) for few-layer phosphorene[19] and our measured surface potential difference between phosphorene and Au.

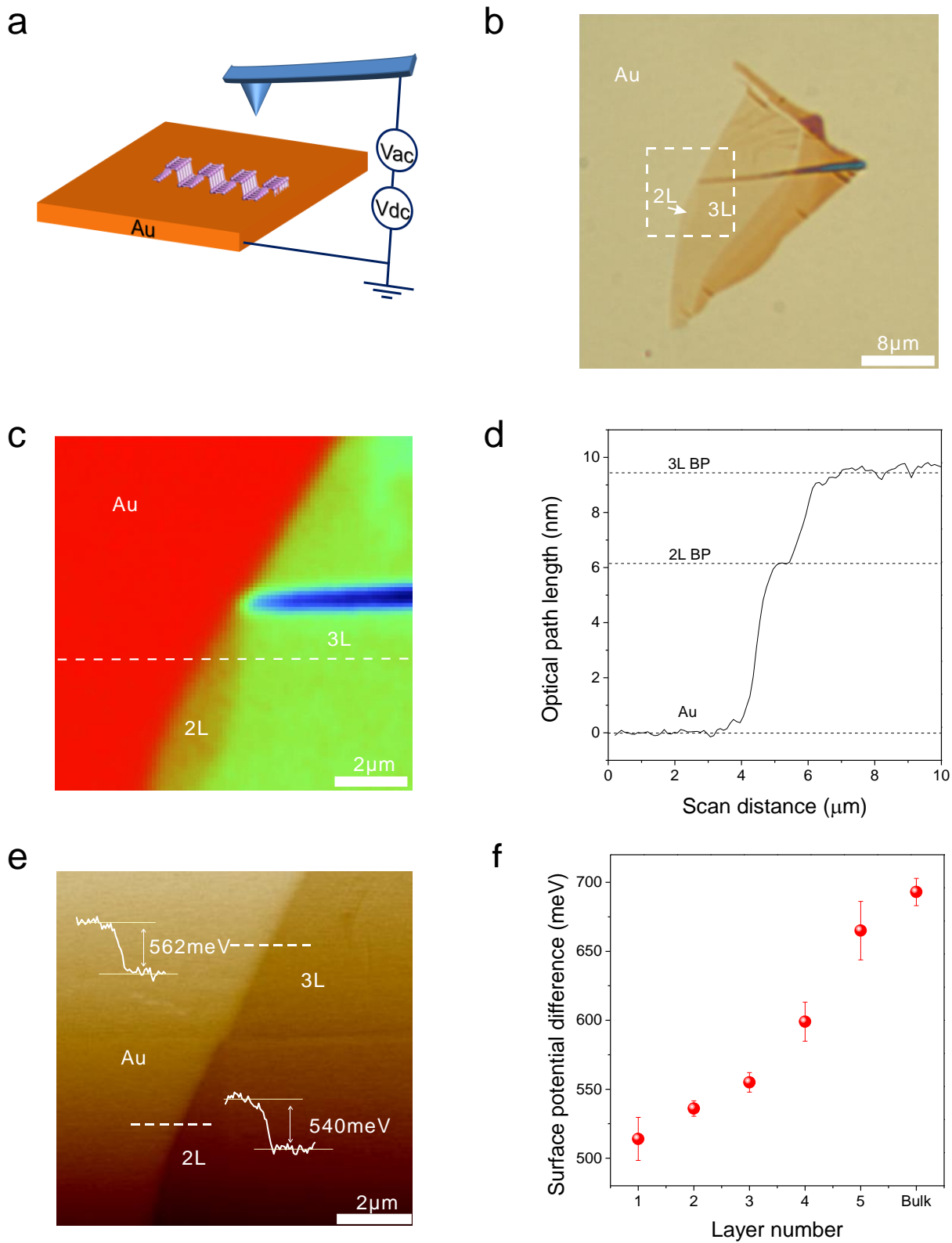

Figure 1

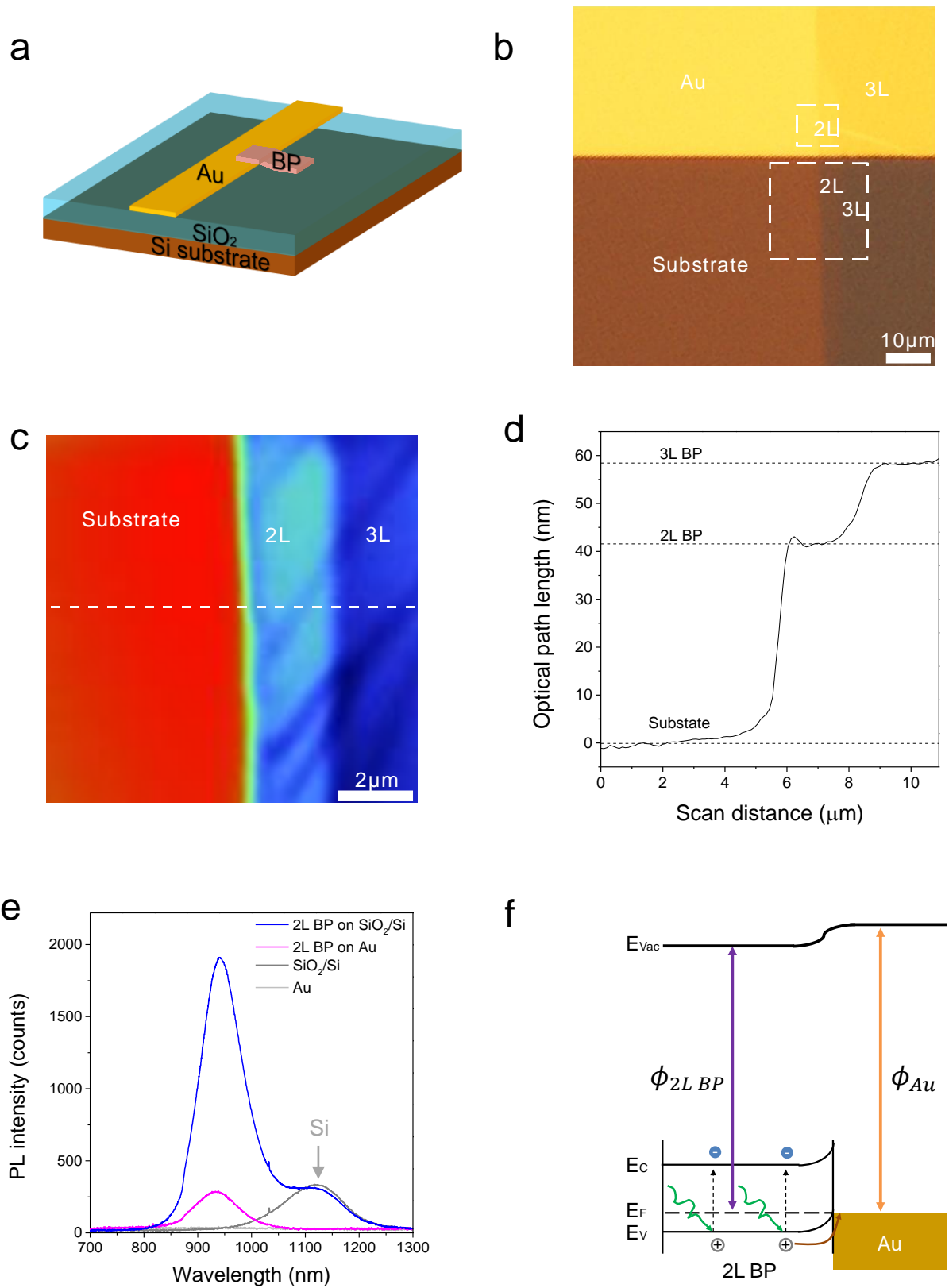

Figure 2

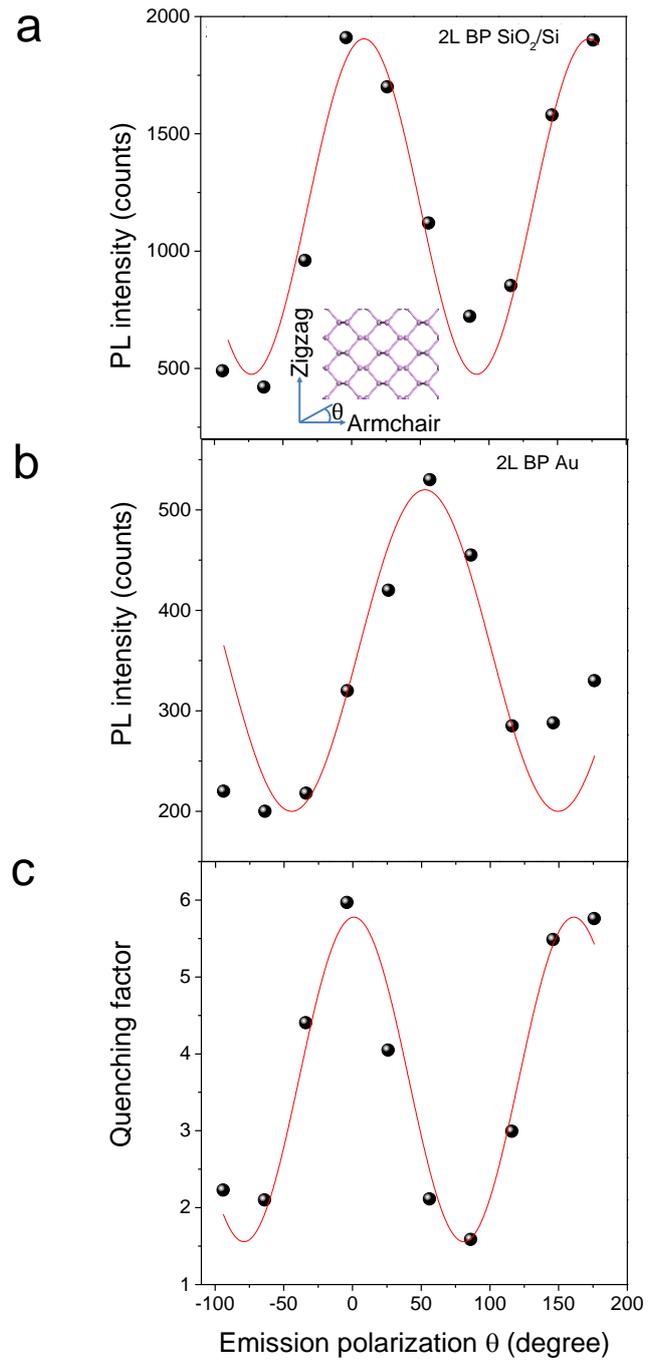

Figure 3

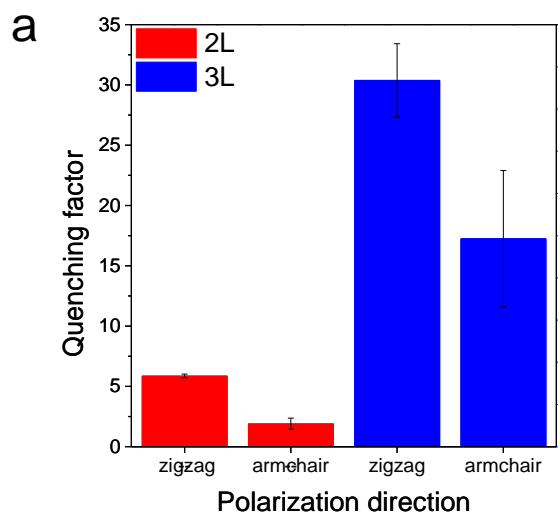 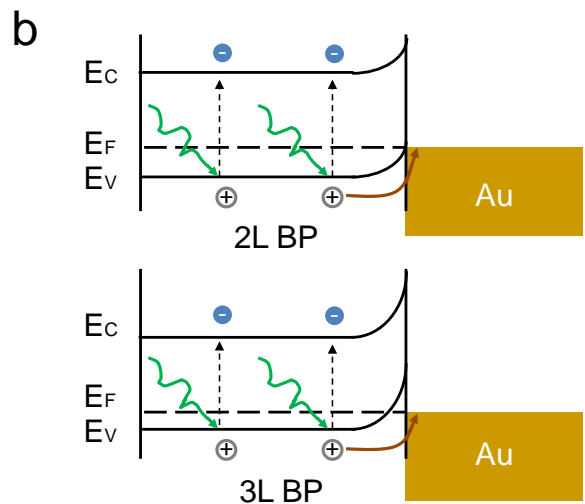

Figure 4

**Supporting Information for**

**Layer-dependent surface potential of phosphorene and anisotropic/layer-dependent charge transfer in phosphorene-gold hybrid system**

Renjing Xu,[1] Jiong Yang,[1] Yi Zhu,[1] Yan Han,[1] Jiajie Pei,[1,2] Ye Win Myint,[1] Shuang Zhang,[1] and Yuerui Lu[1*]

[1]Research School of Engineering, College of Engineering and Computer Science, the Australian National University, Canberra, ACT, 0200, Australia

[2]School of Mechanical Engineering, Beijing Institute of Technology, Beijing, 100081, China

**\*** To whom correspondence should be addressed: Yuerui Lu (yuerui.lu@anu.edu.au)

# 1. Optical path length (OPL) calibration curves for mono- and few-layer phosphorene on SiO₂/Si and gold substrates

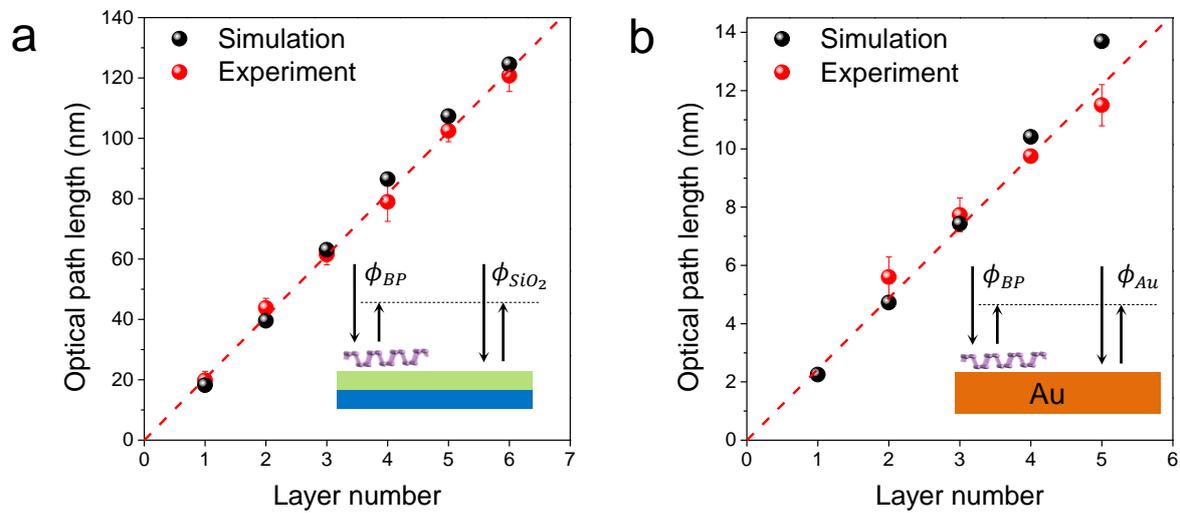

**Figure S1 | OPL calibration curves for mono- and few-layer phosphorene on SiO₂/Si and gold substrates. a**, OPL calibration curves (simulation and experiments) for mono- and few-layer phosphorene on SiO₂/Si (275 nm thermal oxide) substrate. For each layer number of phosphorene, at least five different samples were characterized for the statistical measurements. The red dash line is the linear trend for statistical data measured with the PSI system. Inset is the schematic plot showing the PSI measured phase shifts of the reflected light from the phosphorene flake ($\phi_{BP}$) and the SiO₂/Si substrate ($\phi_{SiO_2}$). **b**, OPL calibration curves (simulation and experiments) for mono- and few-layer phosphorene on gold substrate (monolayer phosphorene only has simulation result). For each layer number of phosphorene, at least two different samples were characterized for the statistical measurements. The red dash line is the linear trend for statistical data measured with the PSI system. Inset is the schematic plot showing the PSI measured phase shifts of the reflected light from the phosphorene flake ($\phi_{BP}$) and the Au substrate ($\phi_{Au}$)

*Numerical Simulation*: Stanford Stratified Structure Solver (S4)[1] was used to calculate the OPL values for mono- and few-layer phosphorene on two different substrates $SiO_2$/Si and gold, respectively (Figure S1). The method numerically solves Maxwell's equations in multiple layers of structured materials by expanding the field in the Fourier-space. In the numerical calculation, the refractive indices used for phosphorene[2], Si, $SiO_2$ and Au were 3.4, 4.15 + 0.04$i$, 1.46 and 0.52 + 2.17$i$, respectively. Both the measured and calculated OPL values for these three semiconductors show a linear relationship with the layer number and they consist well with each other.

2. **Phase-shifting interferometer (PSI) images for phosphorene samples**

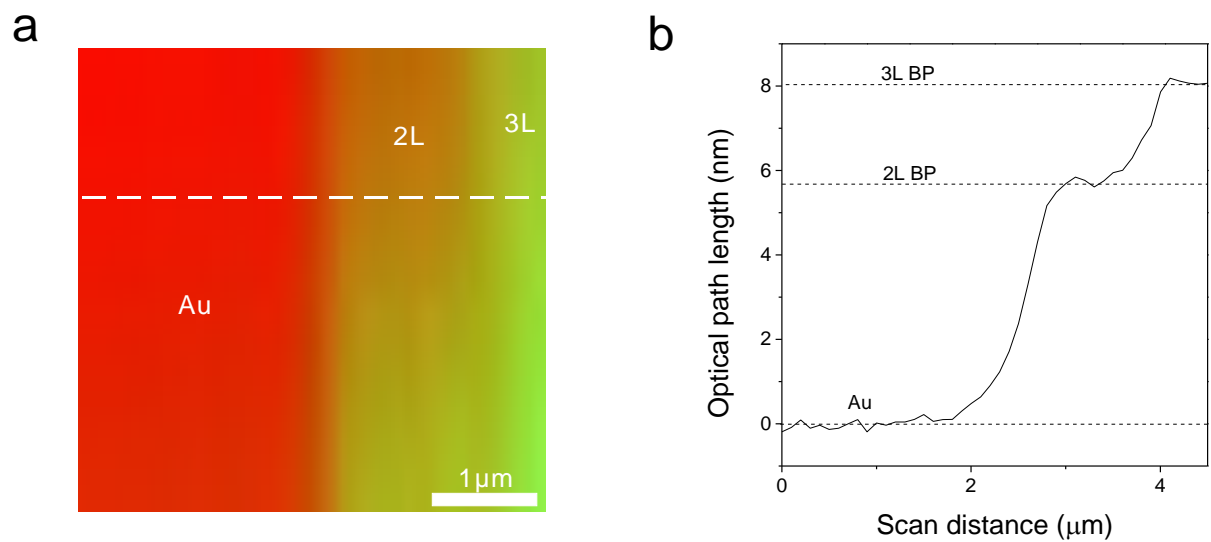

**Figure S2 | PSI images of phosphorene on Au substrate. a**, PSI image of phosphorene samples from the upper box on Au substrate enclosed by the dash line in Figure 2b. **b**, PSI measured OPL values versus scan position for the phosphorene along the dash line in (**a**).

## 3. PSI of another 2L phosphorene sample

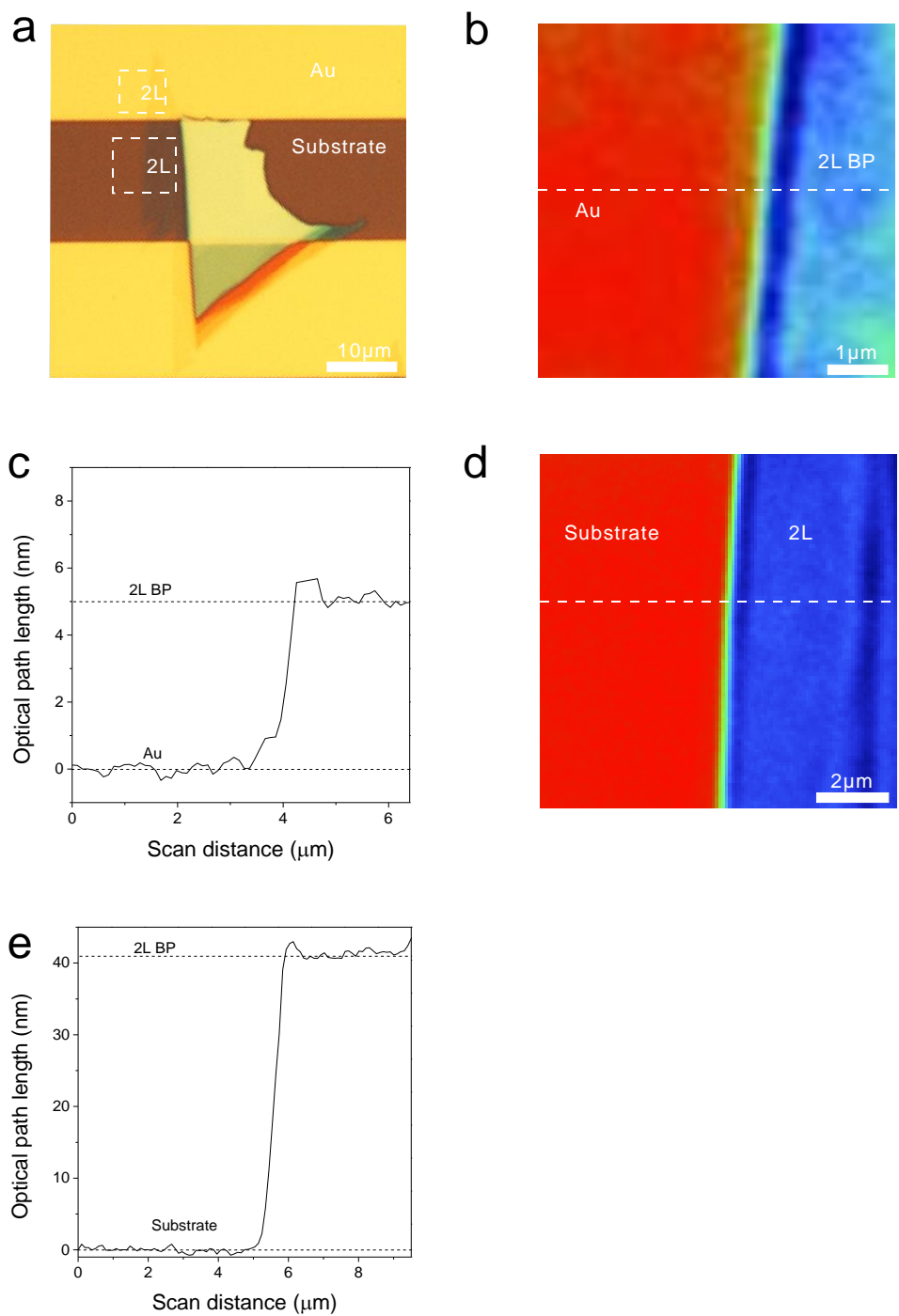

**Figure S3 | PSI of another 2L phosphorene sample. a**, Optical microscope image of the 2L phosphorene sample, with half on SiO$_2$/Si substrate and the other half on Au. **b**, PSI image of 2L phosphorene on Au from the upper box enclosed by the dash line in (**a**). **c**, PSI measured OPL values versus scan position for 2L phosphorene along the dash line in (**b**). **d**, PSI image of 2L

phosphorene on SiO$_2$/Si from the lower box enclosed by the dash line in (**a**). **e**, PSI measured OPL values versus scan position for 2L phosphorene along the dash line in (**d**).

### 4. Kelvin probe force microscope (KPFM) image for monolayer phosphorene sample

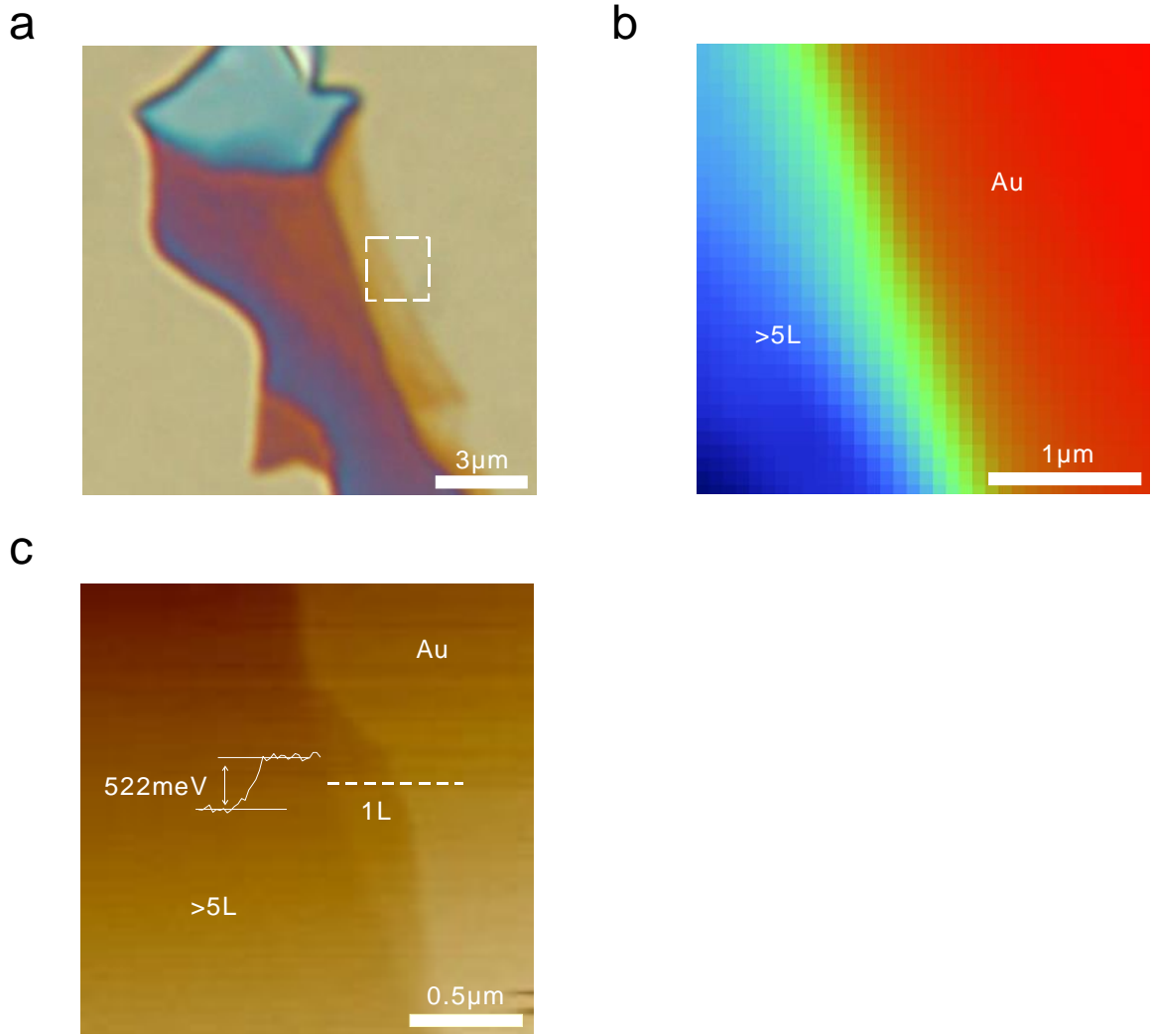

**Figure S4 | KPFM for monolayer phosphorene sample. a**, Optical microscope image of the monolayer phosphorene sample. **b**, PSI image of monolayer phosphorene from the box enclosed by the dash line in (**a**). KPFM image and measurement results of the monolayer phosphorene the box enclosed by the dash line in (**a**).

Here, the monolayer phosphorene is clearly observed in KPFM image as shown in Figure S4c while it is absent under the PSI image in the Figure S4b. This is because the resolution of our PSI system in lateral dimension is ~1 µm, which is limited by optical diffraction. The lateral size of the monolayer phosphorene in Figure S3c is less than 1 µm.